\documentclass{aa}  

\usepackage{graphicx}
\usepackage{txfonts}
\newcommand{\ud}[2]{\mbox{$^{\scriptscriptstyle + #1}_{\scriptscriptstyle - #2}$}}

\begin{document} 

   \title{Thermal X-ray emission identified from the millisecond pulsar PSR J1909-3744}


   \author{N. A. Webb
          \inst{1}
          \and
          D. Leahy \inst{2}
          \and 
          S. Guillot\inst{1,3}
          \and
          N. Baillot d'Etivaux\inst{4}
          \and 
          D. Barret\inst{1}
          \and  
          L. Guillemot\inst{5,6}
          \and 
          J. Margueron\inst{4,7}
          \and 
          M. C. Miller \inst{8}
          }

   \institute{IRAP, Universit\'e de Toulouse, CNRS, CNES, Toulouse, France\\
     \email{Natalie.Webb@irap.omp.eu}
     \and
     Department of Physics \& Astronomy, University of Calgary, Calgary, AB T2N 1N4, Canada 
     \and
     Instituto de Astrof\'isica, Pontificia Universidad Cat\'olica de Chile, Av. Vicu\~{n}a Mackenna 4860, Macul, Santiago, Chile
     \and
     Institut de Physique Nucl\'eaire de Lyon,  CNRS/IN2P3, Universit\'e de Lyon, Universit\'e Claude Bernard Lyon 1, F-69622 Villeurbanne Cedex, France
     \and
     Laboratoire de Physique et Chimie de l’Environnement et de l’Espace, Universit\'e d’Orl\'eans/CNRS, F-45071 Orl\'eans Cedex 02, France 
     \and
     Station de radioastronomie de Nan\c{c}ay, Observatoire de Paris, CNRS/INSU, F-18330 Nan\c{c}ay, France
     \and
     Institute for Nuclear Theory, University of Washington, Seattle, Washington 98195, USA
     \and
     Department of Astronomy and Joint Space-Science Institute, University of Maryland, College Park, MD 20742-2421, USA
   }

   \date{Received ; accepted }

 
  \abstract
   {Pulsating thermal X-ray emission from millisecond pulsars can be used to obtain constraints on the neutron star equation of state, but to date only five such sources have been identified. Of these five millisecond pulsars, only two have well constrained neutron star masses, which improve the determination of the radius via modelling of the X-ray waveform.} 
   {We aim to find other millisecond pulsars that already have well constrained mass and distance measurements that show pulsed thermal X-ray emission in order to obtain tight constraints on the neutron star equation of state.}
   {The millisecond pulsar \object{PSR~J1909--3744} has an accurately determined mass, M = 1.54$\pm$0.03 M$_\odot$ (1 $\sigma$ error) and  distance, D = 1.07$\pm$0.04 kpc. We analysed {\em XMM-Newton} data of this 2.95 ms pulsar to identify the nature of the X-ray emission.}
   {We show that the X-ray emission from \object{PSR~J1909--3744} appears to be dominated by thermal emission from the polar cap. Only a single component model is required to fit the data. The black-body temperature of this emission is  kT=0.26\ud{0.03}{0.02} keV and we find a 0.2--10 keV un-absorbed flux of 1.1 $\times$ 10$^{-14}$ erg cm$^{-2}$ s$^{-1}$ or an un-absorbed luminosity of 1.5 $\times$ 10$^{30}$ erg s$^{-1}$. }
   {Thanks to the previously determined mass and distance constraints of the neutron star \object{PSR~J1909--3744}, and its predominantly thermal emission, deep observations of this object with future X-ray facilities should provide useful constraints on the neutron star equation of state. }



\authorrunning{Webb et al.}
\titlerunning{Thermal X-ray emission from PSR~J1909--3744}

   \maketitle
%

\section{Introduction}
\label{sec:intro}

Fifty years after the discovery of neutron stars, the nature of the
material making up their core remains largely unknown. At densities
above a few times the saturation density of nuclear matter, certain
models predict the existence of exotic components such as hyperons or
unconfined quarks \citep[e.g.][]{latt07}. Neutron stars constitute the
only medium in which nuclei exist in extremely dense but relatively
cold environments. It is therefore essential to understand the
equation of state of this material (meaning its pressure as a
  function of energy density) if we wish to have a full understanding
of the baryonic matter composing the Universe.

Measuring the mass and radius of a neutron star allows us to constrain
the density and pressure of the neutron star. Tight constraints should
be possible with future gravitational wave observations of neutron
star mergers, but to date, only weak constraints are available
\citep[e.g.][]{de18,abbott18}. Using electromagnetic observations,
accurate mass measurements have been made for a number of radio
pulsars, for example, 1.18\ud{0.03}{0.02}~M$_\odot$, for the case of one of
the pulsars in \object{PSR~J1756--2251} \citep{faul05} or
1.4414$\pm0.0002$ M$_\odot$, for \object{PSR~B1913+16} \citep{weis05}.
While there are some accurate mass measurements already available, the
situation is different with radii that are much harder to constrain
\citep[e.g.][]{mill13,mill16b}. A promising way of inferring radii
(and masses) is through modelling the X-ray waveform of a neutron star
with a fairly low magnetic field and which shows predominantly thermal
X-ray emission from its polar caps. This provides a measure of the
mass and radius of the star
\cite[e.g.,][]{pavl97,bogd07,leah11}. Millisecond pulsars have
magnetic fields, $B\lesssim10^9$~G, which are weak enough that
they do not modify the opacity of the neutron star atmosphere
\citep[e.g.][]{hein06}.  Five millisecond pulsars have been shown to
have predominantly thermal X-ray emission and show X-ray
pulsations. These are \object{PSR~J0030+0451},
\object{PSR~J2124--3358}, \object{PSR~J0437--4715},
\object{PSR~J1024--0719} and \object{PSR~1614--2230}
\citep[e.g.,][]{zavl06,bogd08,bogd13,panc12}.  Indeed, loose
constraints have been obtained for the radii of some of these neutron
stars. For example \cite{bogd09} found a lower limit $R>10.4$~km (at
99.9\% confidence) when considering a neutron star of 1.4~M$_\odot$
for \object{PSR~J0030+0451}. Published constraints on the radius for
\object{PSR~J0437--4715} are $R>11.1$~km (99.7\% confidence) for a
neutron star mass of 1.76~M$_\odot$ \citep{bogd13}, but more recent
results suggest a mass of $1.44\pm0.07$~M$_\odot$ \citep{rear16}, which
relaxes the lower limit slightly \citep{guil16}.  However, it is often
difficult to determine accurate radii due partly due to poorly
constrained masses and the degeneracy between the mass and the radius.

Precise knowledge of the mass improves the estimate of the radius, and
well-constrained distance measurements also help somewhat
\citep[e.g.][]{lo13}. This in turn yields a more precise constraint on the
neutron star equation of state.  \object{PSR~J1909--3744} has been
previously detected in the X-ray domain by \cite{karg12}. The 29.7~ks
{\em Chandra} ACIS observations revealed 63 net X-ray counts. The
median energy of these photons is 1.0~keV, which could indicate that
the emission from this pulsar is predominantly soft thermal
emission. In order to determine whether this is indeed the case, we
obtained {\em XMM-Newton} observations of \object{PSR~J1909--3744}. It
is a 2.95 ms pulsar which has both a well-constrained mass of
1.54$\pm$0.03~M$_\odot$ \citep[1 $\sigma$ error, ][]{desv16} and a
very well-constrained distance of $1.07\pm 0.04$~kpc \citep{jone17}
thanks to its proximity and edge-on orientation allowing measurement
of the range and shape of the Shapiro delay \citep{jaco05}.

In this paper we investigate the nature of the X-ray emission with an
aim to constraining the neutron star equation of state. Identifying
new, thermally emitting pulsating millisecond pulsars with excellent
mass and distance constraints will provide new targets for the NICER
mission \citep[e.g.][]{arzo14,oeze16b}, for {\em XMM-Newton}
\citep{jans01}, as well as for future missions, such as {\em Athena}
  \citep{nand13} or {\em Strobe-X} \citep{ray19} all towards finally
  understanding the nature of the material inside neutron stars.

\section{Data reduction and analysis }

\object{PSR~J1909--3744} was observed with {\em XMM-Newton} on 20
September 2016 for a total of 51.6~ks. All three EPIC cameras were
operated in full-frame mode. None of these observations have a high
enough time resolution to detect any X-ray pulsations from this
pulsar. Further information on the observations are given in
Table~\ref{tab:XMMdata}. The optical monitor was not in operation
during these observations.

\begin{table*}
\caption{{\em XMM-Newton} data of \object{PSR~J1909--3744}. The
  cameras and filters used, along with the total exposure time and the
  {\em good time interval} (GTI) used for the analysis are given in ks. The
  radii (source and background) for extracting spectra and light curves
  and the total counts in the source and the background regions are
  also provided.}
\label{tab:XMMdata}      
\centering          
\begin{tabular}{c c c c c c}     
\hline\hline       
Camera & Filter & Exposure & Src radius/ & Src cts/ & Net \\
       &        & (GTI) ks & Bkg radius (\arcsec) & Bkg cts & counts\\
\hline
MOS 1 & Medium & 43.6 (31.6) & 11/60 & 47/407   & 33$\pm$6\\
MOS 2 & Medium & 43.6 (31.4) & 12/60 & 54/391   & 38$\pm$6\\
pn    &   Thin & 50.4 (25.5) & 12/60 & 218/1941 & 140$\pm$12\\
\hline                   
\end{tabular}
\end{table*}

We reduced the raw {\em XMM-Newton} data using the {\em XMM-Newton}
{\tt Science Analysis System} ({\tt SAS}, version 16.1) and the latest
calibration files at the time of the data reduction (CCFs, August
2017). The MOS data were reduced using the {\tt SAS} task 'emproc'
and the {\tt SAS} task 'barycen' was used to barycentre the data,
using the coordinates of the pulsar.  The event lists were filtered
with the \#XMMEA\_EM flag, and zero to 12 of the pre-defined patterns
(single, double, triple, and quadruple pixel events) were retained. We
identified periods of high background in the same way as described in
the {\em XMM-Newton} {\tt SAS}
threads\footnote{http://xmm.esac.esa.int/sas/current/documentation/threads/}
and the good time interval is given in parentheses in
Table~\ref{tab:XMMdata}. We also filtered in energy, using the range
0.2--12.0 keV.  The pn data were reduced using the 'epproc'
and zero to 4 of the pre-defined patterns (single and double events) were
retained, as these have the best energy calibration. Again we used the
task 'barycen' to barycentre the data.  The background was treated
in the same way as for the MOS data. We used the \#XMMEA\_EP filtering
and the same energy range as for the MOS.

\begin{figure}
  \centering
  \includegraphics[width=9cm, angle=0]{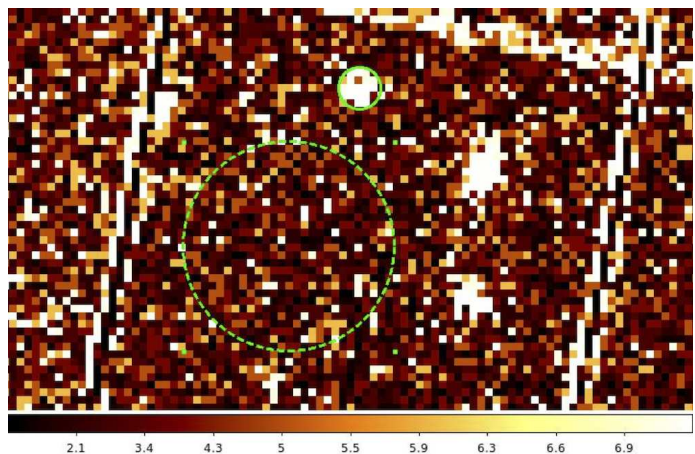}
  \caption{pn image of the region around \object{PSR~J1909--3744}.
    The region is shown with a solid circle (green in the colour version,
    12\arcsec\ radius) and the background region, shown with a dashed
    line (60\arcsec\ radius). The circles show the radii used for the
    extraction regions for the pn data. The lighter the colour, the
    greater the number of counts.}
  \label{fig:image1909}
\end{figure}

The {\tt SAS} provides a task ('especget'), which allows the user
to find an extraction region that optimises the source signal with
respect to the background. We extracted the data using 'especget'
and the regions used are given in Table~\ref{tab:XMMdata}. The
background was chosen from a source-free region close to the source
and on the same CCD.  These regions can be seen in 
Figure~\ref{fig:image1909}. To create the spectra, we re-binned the data
into 5eV bins as recommended in the SAS threads.  We used the {\tt
  SAS} tasks 'rmfgen' and 'arfgen' to generate a
`re-distribution matrix file' and an `ancillary response file', for
each spectrum. The pn data were binned to contain at least ten counts
per bin and the MOS data to contain five counts per bin.  The spectra
were fitted using {\em Xspec} version 12.5 \citep{arna96}.

The light curves were extracted using the same regions as for the
spectra and using the maximum temporal resolution of the camera in
use, as well as using the same temporal range for the source and
background light curves for each camera. The source light curves were
corrected for the background using the task 'epiclccorr'.

\section{Results}

\begin{figure}
 \includegraphics[width=9.3cm, angle=0]{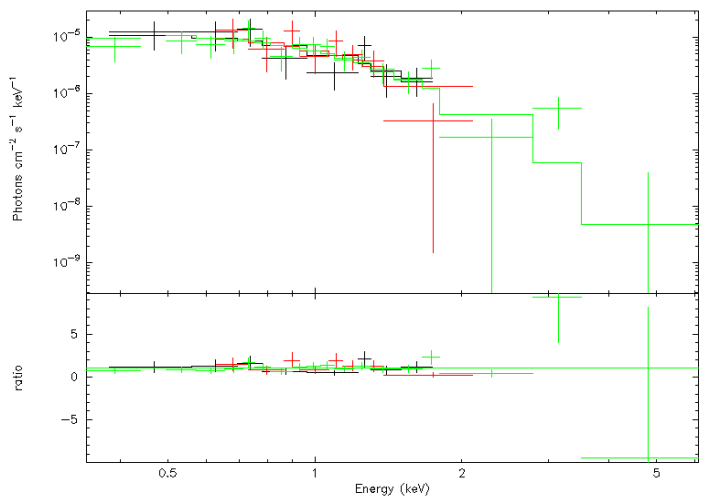}
  \caption{Spectrum of \object{PSR~J1909--3744}. The data is represented by crosses from MOS1 (black), MOS2 (red) and pn
    (green) of \object{PSR~J1909--3744}. The top plot shows the data fitted with fixed absorption
    (tbabs) and a hydrogen atmosphere model ({\tt nsatmos}, the *
    model in Table~\ref{tab:SpecFit}, green solid line). The bottom plot shows the 
    same data divided by the model and shown as residuals, in
      units $\sigma$, which is the error bar of each data point.}
  \label{fig:1909spec}
\end{figure}

\begin{figure}
  \centering
 \includegraphics[width=\columnwidth, angle=0]{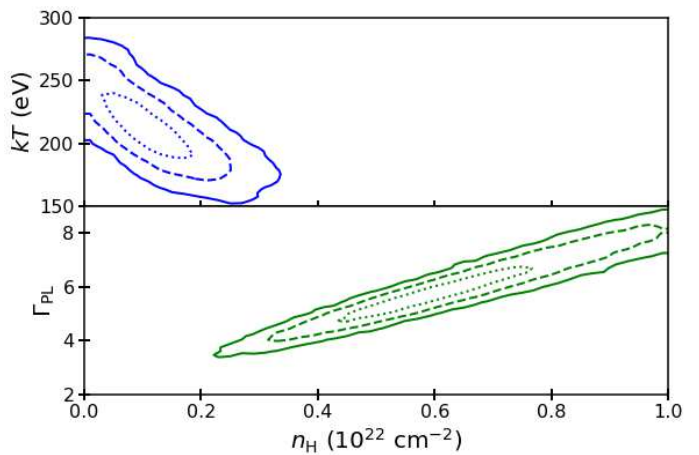}
  \caption{1, 2, and 3 $\sigma$ confidence contours. The top plot shows
    the degeneracy between the $n_{\rm H}$ and the temperature of the
    black body for \object{PSR~J1909--3744}. On the bottom plot 1, 2, and 3
    $\sigma$ confidence contours show the degeneracy between the
    $n_{\rm H}$ and the photon index $\Gamma_{\rm PL}$ of the
    power-law fit.}
  \label{fig:contours}
\end{figure}

\object{PSR~J1909--3744} was significantly detected with all three
cameras, with 140$\pm$12, 33$\pm6$, and 38$\pm6$ background subtracted
counts detected with the pn, the MOS 1, and the MOS2 respectively, see
also Table~\ref{tab:XMMdata}. To determine if the X-ray emission from
\object{PSR~J1909--3744} is thermal or non-thermal, we fitted simple
models to the MOS and pn spectra simultaneously. We chose a black-body
model as a proxy for thermal emission and a power-law model to
represent non-thermal emission. We also included a model for the
absorption due to the interstellar medium, namely 'tbabs' in 
  Xspec with the \cite{wilm00} abundances. As the data were binned to
contain only five to ten counts per bin, we used the C-statistic to assess
the accuracy of the fit. Table~\ref{tab:SpecFit} gives the results of
the spectral fitting. The absorption is not constrained when using an
absorbed black-body model. However, no absorption is not
physical. Different H {\sc I} maps indicate that we could expect
$n_{\rm H}\sim6.5\times10^{20}$~cm$^{-2}$ \citep{kalb05} to
$\sim7.5\times10^{20}$~cm$^{-2}$ \citep{dick90}. Alternatively, using
the conversion of dispersion measure to $n_{\rm H}$ \citep{he13} and
the dispersion measure of 10.4~pc~cm$^{-3}$ determined by
\cite{jone17}, we expect $n_{\rm H}\sim3.1\times10^{20}$~cm$^{-2}$. We
therefore used this value to fit the spectra and the model fits are
also given in Table~\ref{tab:SpecFit}.  We explore the degeneracy of
the absorption and the temperature of the black body in
Figure~\ref{fig:contours} (top panel).

As can be seen in Table~\ref{tab:SpecFit}, the power law with fixed
absorption gave a poor fit (C-statistic of 52.5 as opposed to 31.4 for
a black-body model with a fixed absorption component, see
Table~\ref{tab:SpecFit}).  However, the C-statistic does not allow us
to directly determine the accuracy of the fit. We therefore did this in
'Xspec' by simulating spectra based on the model 100,000 times,
using parameter values drawn from a Gaussian distribution centered on
the best fit. We then fitted each fake data set and calculated the test
statistic. This goodness-of-fit testing only allows us to reject a
model with a certain level of confidence (i.e., it can not provide a
probability that the model is correct). Doing this for the fixed
absorption and the black body, the model is rejected only at the
10.63\% level, whereas the fixed absorption and the power law is
rejected at 54.98\%. This is not very conclusive, so in
order to use $\chi^2$ statistics we also tried
fitting the data that was binned to contain a minimum of 20 counts per bin. The derived parameters we obtained
when fitting the binned data are very similar to those determined when
using the fixed absorption and either the black-body or the power-law
models. For the fixed absorption and the black-body model we find a
$\chi^{\scriptscriptstyle 2}_{\scriptscriptstyle \nu}$ = 0.67, with seven
degrees of freedom and a null hypothesis that the observed data are
drawn from the model of 0.7. For the fixed absorption and the
power-law model we find a $\chi^{\scriptscriptstyle
  2}_{\scriptscriptstyle \nu}$ = 1.98, with seven degrees of freedom and a
null hypothesis that the observed data are drawn from the model of
0.05. This further reinforces the notion that the data are better
fitted using a black-body model and that the emission is therefore
predominantly thermal, with no need for a harder power-law tail as is
sometimes seen in pulsar spectra, especially with low signal-to-noise ratio. Further, as can be seen in Figure~\ref{fig:contours} (bottom
panel), fitting the data with a power law requires a high photon index
(reminiscent of a thermal spectrum) at the same time as a high
absorption, much higher than the expected absorption. Both indicate
that the power-law model is unlikely to describe the data.

\begin{figure}
  \centering
  \includegraphics[width=\columnwidth, angle=0]{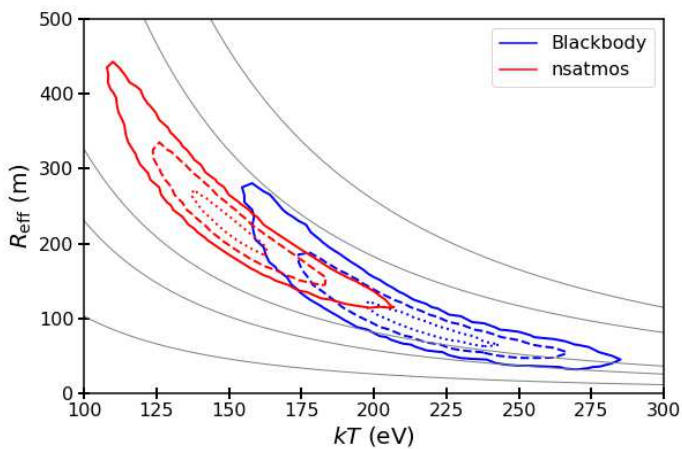}
  \caption{1, 2, and 3 $\sigma$ confidence contours from fitting
    the X-ray spectrum of \object{PSR~J1909--3744}. This shows the
    degeneracy between the temperature and size of the emitting
    surface when using the black-body model and the {\tt nsatmos}
      model.  For the latter, the model with fixed neutron star radius
      ($R=10.6$ km, model $^{z}$ in Table~\ref{tab:SpecFit}) was used.
      Lines of constant (un-absorbed) bolometric flux are shown in
      grey, ranging from $10^{-15}$ to
      $10^{-13}$~erg~s$^{-1}$~cm$^{-2}$, in steps of 0.5 dex.
  }
  \label{fig:RadiusContours}%
\end{figure}

We estimate the radius of the emitting region in a crude way by taking
the emission over the whole observation, in order to have a spectrum
with enough of a signal-to-noise ratio to fit with a model, and therefore smear
out the effects of the projected spot area with rotation phase.
Without the detection of pulsations, we also do not know if the
emission is coming from one or two spots, so the emission radius is
only an estimate and does not take into account that the spots are
likely to be fairly flat, in addition to other factors outlined in
\cite{szar17}. We use the {\tt bbodyrad} model in {\em Xspec} to
estimate an approximate emitting radius of $\sim$58~m (see
Table~\ref{tab:SpecFit}) typical of emission from polar caps with low
signal-to-noise spectra \citep[e.g.][and see
  Section~\ref{sec:disc}]{bogd06}. The emission is therefore coming
from at least one polar cap. Given the low signal-to-noise, the
degeneracy between the emitting area
and the temperature of the emission is given in
Figure~\ref{fig:RadiusContours} for the black-body model. The same
plot for the neutron star atmosphere model {\tt nsatmos} is also shown in
that same figure.  Contours in Figures~\ref{fig:contours} and
  \ref{fig:RadiusContours} were obtained by sampling the parameter
  spaces via Markov Chain Monte Carlo in \emph{Xspec} (command {\tt
    chain} with the Goodman-Weare algorithm, 200 walkers, 100,000
  steps, and a 20\% burn-in).  The un-absorbed luminosity of this MSP
is $1.5\times10^{30}$~erg~s$^{-1}$ (0.2--10.0 keV), similar to other
thermally emitting MSPs.

\begin{table*}
\caption{Results from fitting different models to EPIC spectra of
  \object{PSR~J1909--3744}. Column one gives the interstellar absorption
  ($\times10^{22}$~atom~cm$^{-2}$); columns two and three give the black-body temperature or the
  power-law index ($\Gamma$) respectively; column four gives the normalisation of the model
  (radius$^2$(km)/distance$^2$(10~kpc) for the black-body model or
  fraction of the neutron star surface emitting for the {\tt nsatmos}
  model); column five gives the radius of the emitting region in the case of the thermal
  models and for the {\tt nsatmos} models assuming a neutron star
  radius of 10.6~km, or photons~$^{-1}$~keV$^{-1}$~cm$^{-2}$~s$^{-1}$
  at 1~keV for the power-law model and the goodness of fit measured
  using the C-statistic; column six gives the number of degrees of freedom; columns seven and eight give an estimate of the un-absorbed and absorbed
  fluxes in the 0.2--10.0~keV band
  ($\times10^{-14}$~erg~cm$^{-2}$~s$^{-1}$), respectively. All the errors are given
  for 90\% confidence for one interesting parameter. For the
    fluxes, the errors are for 68\% confidence. No error bars on a
  fittable parameter implies that the value for that parameter was
  fixed to the displayed value.}
\label{tab:SpecFit}      
\begin{center}          
\begin{tabular}{c c c c c c c c }     
\hline\hline       
$n_{\rm H}$ & kT (keV) & $\Gamma$ & Norm.  & Radius (m) & C-stat (dof) & Ab. flux & Unab. flux\\
\hline

0.03                & 0.26\ud{0.03}{0.02} &                  & 0.28\ud{0.2}{0.1} & 57\ud{2}{1}   & 31.4 (32) & 1.04\ud{0.06}{0.15} &  1.22\ud{0.07}{0.17} \\
0.03                &                     & 2.1\ud{0.2}{0.3} & 46\ud{6}{6}       &               & 52.5 (32) & 1.66\ud{0.22}{0.17} &  2.13\ud{0.25}{0.14} \\
0.03\ud{0.14}{0.03} & 0.25\ud{0.04}{0.05} &                  & 0.3\ud{0.9}{0.1}  & 59\ud{58}{11} & 31.3 (31) & 1.02\ud{0.02}{0.80} &  1.31\ud{0.13}{0.76} \\
0.48\ud{0.30}{0.30} &                     & 4.8\ud{1.8}{1.3} & 210\ud{176}{127}  &               & 30.9 (31) & 1.04\ud{0.03}{0.17} &  155\ud{565}{128} \\
0.03$^*$            & 0.15\ud{0.14}{0.02} &                  & 0.00021\ud{0.00073}{0.00018} & 153\ud{172}{95} & 32.3 (31)  & 1.10\ud{0.08}{0.11} & 1.31\ud{0.12}{0.10} \\
0.03$^z$            & 0.18\ud{0.03}{0.03} &                  & 0.00033\ud{0.00040}{0.00018} & 192\ud{94}{62}  & 32.6 (32)  & 1.10\ud{0.08}{0.10} & 1.31\ud{0.11}{0.09} \\ 
0.03$^y$            & 0.07\ud{0.01}{0.01} &                  & 0.075                        & 2.9             & 171.6 (33) & 0.76\ud{0.06}{0.12} & 1.33\ud{0.11}{0.17} \\
\hline
\end{tabular}
\end{center}
$^*$ - {\tt nsatmos} model, mass = 1.54 M$_\odot$, distance = 1.07 kpc and 7.4 $<$ radius $<$ 30$^a$ km (90\% range)\\
$^z$ - {\tt nsatmos} model, mass = 1.54 M$_\odot$, distance = 1.07 kpc and radius = 10.6 km (see Sect.~\ref{sec:disc})\\
$^y$ - {\tt nsatmos} model, mass = 1.54 M$_\odot$, distance = 1.07 kpc and radius = 10.6 km and normalisation of 0.075 (the area corresponding to the classical radius of this neutron star, see Sect.~\ref{sec:disc})\\
$^a$ - reached the limit of model parameter in {\em Xspec}
\end{table*}

Fitting the X-ray spectrum with a neutron star atmosphere model ({\tt
  nsatmos} in {\em Xspec}, \citealt{hein06}), we obtain an equally
accurate fit, as with the black body when we use the known distance and
mass of the neutron star given in Section~\ref{sec:intro} (see
Table~\ref{tab:SpecFit}). The data fitted with this spectral model can
be seen in Figure~\ref{fig:1909spec}.

Whilst the best time resolution (73.4~ms for the pn in full-frame
mode) is too coarse to detect any X-ray pulsations of the pulsar
(2.95~ms), and the length of the observation, $\sim$9~h, is also too
short to detect the binary orbital period of 1.5~d \citep{jaco05}, we
checked the background subtracted light curve for variability using a
Kolmogorov-Smirnov test and a $\chi^2$ probability of constancy
test. Both tests gave a probability of constancy of between 0.1 and
0.01 (depending on the camera and the test), confirming that the X-ray
light curves show no evidence for variability.

\section{Discussion}
\label{sec:disc}

Analysis of the X-ray spectrum shows that the X-ray emission from
\object{PSR~J1909--3744} appears to be predominantly thermal. This
millisecond pulsar has a very tight mass constraint, similar to that
of \object{PSR~J1614-2230} which has a mass contraint of
1.928$\pm$0.017 M$_\odot$ \citep[1 $\sigma$,][]{fons16}. It also
exhibits predominantly soft X-ray emission, making it a useful
candidate for future studies to constrain the neutron star equation of
state, especially if X-ray pulsations can be detected (see
Section~\ref{sec:intro}). However, the current short observations
result in signal-to-noise ratios too low to make any such constraints.

The radius estimate of the X-ray emitting area is small compared to
the size of a neutron star, which typically has a radius of
10--14~km. The emission is therefore likely to be coming from the
polar caps of \object{PSR~J1909--3744}, as is expected from such old
objects where the surface of the neutron star has cooled so that it is
no longer detectable in the X-ray band. The emitting radius is
somewhat smaller than the classical radius of a polar cap, $R_{pc}=
\sqrt{\frac{2 \pi R}{c P}} R$ \citep[e.g.][]{derm94}, where $R$ is the
neutron star radius, $c$ the speed of light in a vacuum and $P$ the
rotation period of the neutron star. The latter is $\sim2.9$~km if we
suppose a 10.6~km radius for the \object{PSR~J1909--3744} neutron
star, the average of the best radius values for a neutron star of
$M=1.5$~M$_\odot$ as determined by \cite{oeze16} who undertook a
comprehensive study of 12 neutron stars.  Considering the range of
typical radii of neutron stars (10--14 km), the classical polar cap
radius is between 2.7 and 4.4 km. \cite{bogd06} state that the radius
determined through fitting the pulsar spectrum can be smaller than
expected when the spectrum has been fitted with a single-temperature
model, but that a higher signal-to-noise spectrum would require a
two-temperature model, as for \object{PSR~J0030+0451},
\object{PSR~J2124--3358} \citep[e.g.][]{bogd08}, and
\object{PSR~J0437--4715} \citep{bogd13}. Fitting the spectrum with two
black bodies does not improve the fit, again probably due to the low
signal-to-noise ratio. Further, it is known that the radii of neutron stars
are underestimated when using a black-body model as opposed to a
realistic neutron star atmosphere model
\citep[e.g.][]{hein06}. However, assuming the classical radius of the
polar cap for this neutron star (2.9 km) in the {\tt nsatmos} model
provides a poor fit to the data with a C-statistic=171.6, 33 degrees
of freedom, suggesting that such a radius is too large. In fact,
  the {\tt nsatmos} model produces a polar cap with size $\sim$
  100--400~m, slightly larger than the $\sim$ 50--300~m radius
  obtained with the black-body model (see
  Figure~\ref{fig:RadiusContours}).

The energy loss rate due to spin down was corrected for the Shklovskii effect,
\.E, is $4.3\times10^{33}$~erg~s$^{-1}$ \citep{smit17}. This value is
similar to the values of \.E obtained for the majority of the
thermally emitting MSPs in 47~Tuc \citep{bogd06} and in other thermally
emitting MSPs \citep{abdo13}, which have also been detected in
gamma-rays. It is therefore not surprising that this pulsar has just
joined the list of {\em Fermi LAT} detected pulsars
\citep{smit17}. Two gamma-ray peaks are observed, along with a very
narrow radio peak \citep{smit17}. The X-ray luminosity ($L_{\rm X}$)
to \.E ratio for \object{PSR~J1909--3744} is similar to other
thermally emitting millisecond pulsars, $4.4\times10^{-4}$. The 47~Tuc
pulsars have ratios between $\sim10^{-4} - 10^{-3}$ \citep{bogd06} and
the non-globular cluster, thermally emitting pulsars have values of
$\sim10^{-4}$ \cite{abdo13,mare11}. Modelling of the gamma-ray and
radio light curves should eventually reveal further constraints on the
pulsar geometry \citep[e.g.][]{john14}.

The magnetic field at the light cylinder ($r_c$) can also be
calculated if we assume a simple dipole model for the neutron star,
where $r_c = cP/2\pi$.  We determine $7.4\times10^4$~G, which is again
similar to values determined for the thermally emitting MSPs in 47~Tuc
\citep{bogd06} and average for {\em Fermi} LAT detected pulsars
\citep{abdo13}.

\section{Concluding remarks}

To obtain an accurate constraint on the radius of the neutron star
\object{PSR~J1909--3744}, longer and higher time resolution
observations are required to search for X-ray pulsations. If such
pulsations are detected, the excellent mass and distance constraints
combined with modelling of the X-ray light curve and spectra will allow
strong constraints to be made on the neutron star equation of
state. Whilst this source is likely to be too faint for {\em NICER} given that
it is a factor of two fainter than \object{PSR~J1614--2230} and that there
are sources at least as bright as \object{PSR~J1909--3744} at only
~1\arcmin\ which would cause a high background for {\em NICER}
observations, very long observations with {\em XMM-Newton} or shorter
{\em Athena} observations will be able to achieve good enough quality
spectra and light curves to constrain the radius to a few percent.

\begin{acknowledgements}
NAW, DB, and SG acknowledge the CNES for their support of this work. SG
also acknowledges the support of FONDECYT through the Postdoctoral
Project 3150428. DL was supported in part by the Natural Sciences and
Engineering Research Council of Canada and in part by the Observatoire
Midi-Pyr\'en\'ees. We thank the anonymous referee for raising many
valid points. The work in this paper is based on observations obtained
with {\em XMM-Newton}, an ESA science mission with instruments and
contributions directly funded by ESA Member States and NASA.
\end{acknowledgements}

%
   \bibliographystyle{aa} 
  \bibliography{PSRJ1909v6.bib} 
%

\end{document}